\documentclass[sigconf]{acmart}

\usepackage{booktabs} 
\usepackage{color, colortbl}
\usepackage{booktabs} 
\usepackage{graphicx}
\usepackage{courier}
\usepackage{makecell}
\usepackage{float}
\usepackage{bm}
\usepackage{multirow}
\usepackage[flushleft]{threeparttable}
\usepackage{flushend}
\usepackage{bbm}
\usepackage{mathrsfs}
\usepackage{enumitem}
\usepackage[ruled, linesnumbered]{algorithm2e} 
\newcommand{\ignore}[1]{}
\newcommand\ChangeRT[1]{\noalign{\hrule height #1}}
\newcommand{\ie}{\emph{i.e., }}

\newcommand{\eg}{\emph{e.g., }}

\usepackage{etoolbox}

\copyrightyear{2020} 
\acmYear{2020} 
\setcopyright{acmcopyright}\acmConference[SIGIR '20]{Proceedings of the 43rd International ACM SIGIR Conference on Research and Development in Information Retrieval}{July 25--30, 2020}{China}
\acmBooktitle{Proceedings of the 43rd International ACM SIGIR Conference on Research and Development in Information Retrieval (SIGIR '20), July 25--30, 2020, China}
\acmPrice{15.00}
\acmDOI{10.1145/3397271.3401181}
\acmISBN{978-1-4503-8016-4/20/07}

\begin{document}
\fancyhead{}
\title{Neural Interactive Collaborative Filtering}
\author{Lixin Zou$^{1}$, Long Xia$^{2}$, Yulong Gu$^{3}$, \\
Xiangyu Zhao$^{4}$, Weidong Liu$^{1}$, Jimmy Xiangji Huang$^{2}$, Dawei Yin$^{5}$}

\affiliation{
  \institution{$^{1}$Tsinghua University, China, $^{2}$York University, Canada \\
 $^{3}$JD.com, China, $^{4}$Michigan State University, USA, $^{5}$Baidu Inc., China}
}
\email{{zoulx15,liuwd}@mails.tsinghua.edu.cn, {longxia,jhuang}@yorku.ca}
\email{guyulongcs@gmail.com, zhaoxi35@msu.edu, yindawei@acm.org}







\begin{abstract}
In this paper, we study collaborative filtering in an interactive setting, in which the recommender agents iterate between making recommendations and updating the user profile based on the interactive feedback. The most challenging problem in this scenario is how to suggest items when the user profile has not been well established, \ie recommend for cold-start users or warm-start users with taste drifting. Existing approaches either rely on overly pessimistic linear exploration strategy or adopt meta-learning based algorithms in a full exploitation way. In this work, to quickly catch up with the user's interests, we propose to represent the exploration policy with a neural network and directly learn it from the feedback data. Specifically, the exploration policy is encoded in the weights of multi-channel stacked self-attention neural networks and trained with efficient Q-learning by maximizing users' overall satisfaction in the recommender systems. The key insight is that the satisfied recommendations triggered by the exploration recommendation can be viewed as the exploration bonus (delayed reward) for its contribution on improving the quality of the user profile. Therefore, the proposed exploration policy, to balance between learning the user profile and making accurate recommendations, can be directly optimized by maximizing users' long-term satisfaction with reinforcement learning. Extensive experiments and analysis conducted on three benchmark collaborative filtering datasets have demonstrated the advantage of our method over state-of-the-art methods.

\end{abstract}
\maketitle
{\fontsize{8pt}{8pt} \selectfont
\textbf{ACM Reference Format:}\
Lixin Zou, Long Xia, Yulong Gu, Xiangyu Zhao, Weidong Liu, Jimmy Xiangji Huang, Dawei Yin. 2020. Neural Interactive Collaborative Filtering. In \text{\it Proceedings of the 43rd International ACM SIGIR} {\it Conference on Research and Development in Information Retrieval (SIGIR '20), July 25--30, 2020, Virtual Event, China.} ACM, NY, NY, USA, 10 pages. https://doi.org/10.1145/3397271.3401181 }

\section{Introduction}
Over the past decade, recommender systems have shown great effectiveness and become an integral part of our daily lives. Recommendation by nature is an interactive process: a recommender agent suggests items, based on the user profile; users provide feedback on the suggested items; the agent updates the user profile and makes further recommendations. This kind of interactive recommendation paradigm has been widely deployed in real-world systems~(\eg personalized music recommendation in Spotify\footnote{https://www.spotify.com/}, product recommendation in Amazon\footnote{https://www.amazon.com/}, image recommendation in Pinterests\footnote{https://www.pinterest.com/}) and has attracted a lot of interest from the research community~\cite{steck2015interactive,zou2020dynaq}. 

A key challenge in the interactive recommendation is to suggest items with insufficient observations, especially for interactive collaborative filtering where there is no content data to represent users and items and the only observations are users' ratings~\cite{zhao2013interactive}. It poses a ``chicken-or-the-egg'' problem in providing accurate recommendations since satisfied recommendations require adequate observations of user's preferences. Besides, it is inevitable because we only have partial observations or even no observations for the cold-start users or warm-start users with taste drifting, which constitute the main user group. Therefore, a persistent and critical problem in interactive collaborative filtering is how to quickly capture user's interests while not compromising his/her recommendation experience, \ie how to balance between the goals of learning the user profile (\ie exploration) and making accurate recommendations (\ie exploitation)?

The existing approaches mainly studied this problem in two directions: {\bf (1)} {\bf MAB} (\textbf{m}ulti-\textbf{a}rmed \textbf{b}andits) approaches and {\bf (2)} {\bf Meta-Learning} approaches. {\bf (1)} The {\bf MAB} approaches formulate the problem as multi-armed bandits or contextual bandits, and solve it with intricate exploration strategies, such as GLM-UCB and Thompson Sampling~\cite{li2010contextual,chapelle2011empirical,zhao2013interactive}. However, to achieve provably low bounds, these approaches optimize the recommendations in the worst case and result in overly pessimistic recommendations that may not be able to achieve the overall optimal performance. Additionally, these methods are usually computationally intractable for non-linear models, which terrifically limits its usage in recent advanced deep models~\cite{cheng2016wide,he2017neural}.
{\bf (2)} Recently, {\bf meta-learning} approaches, which can fast adapt model on newly encountered tasks, have been leveraged to solve the cold-start recommendation. Existing methods treat suggesting items for different users as different tasks and aim to learn a learning algorithm that can quickly identify user preferences after observing a small set of recommendations, \ie the support set. The meta-learning perspective is appealing since it avoids the complexity of hand-designing sophisticated exploration policies and enables us to take advantage of deep neural networks. However, these approaches ignore the performance on the support set, which may lead to the recommendation of highly irrelevant items and terrible user experience at the phase of constructing the support set. Even worse, these methods perform lousy when faced with users' tastes drifting or poor quality support set due to its deficiency in actively exploring users' interests and excessive dependence on the heuristically selected support set. 

Rather than hand-designing the sophisticated exploration policies, we propose a framework named neural interactive collaborative filtering~(NICF), which regards interactive collaborative filtering as a meta-learning problem and attempts to learn a neural exploration policy that can adaptively select the recommendation with the goal of balance exploration and exploitation for different users. In our method, the exploration policy is structured as a sequential neural network, which consists of two parts. The first part embeds the user profile by feeding past recommendations and user's feedback into multi-channel stacked self-attention blocks to separately capture the information of versatile user feedback. The second part, the policy layer, generates the recommendation with a multi-layer perceptron. Therefore, the sequential neural network can update the user profile based on the historical recommendations and the exploration policy is encoded in the weights of the neural network. In this work, we propose to directly optimize the weights of exploration policy by maximizing the overall users' satisfaction throughout the recommendation journey with an efficient reinforcement learning (RL) algorithm. It is meaningful in two aspects: {\bf (1)} The ultimate goal of exploration/exploitation is to maximize users' overall engagement during the interactive recommendation. {\bf (2)} From the perspective of reinforcement learning, it is insightful since the satisfied recommendations triggered by an exploration recommendation can be viewed as the exploration bonus (delayed reward) for its contribution on improving the quality of the user profile. Therefore, optimizing the sum of immediate rewards and delayed rewards can be viewed as maximizing the balance between the rewards for providing accurate personalized recommendations and the rewards for exploring user's interests, which can be effectively solved by RL. By doing so, the learned exploration policy thus can act as the learning process for interaction recommendations and constantly adapt its strategy when deployed with cold-start or warm-start recommendation (analyzed in Section \ref{sec:case}).

The NICF exhibits following desirable features: {\bf (1)} It avoids the overly pessimism and complexity of existing hand-designing exploration policies for interactive collaborative filtering. {\bf (2)} It can be incorporated with any advanced deep model for recommendations~\cite{wang2015collaborative,cheng2016wide}, which can capture much more non-linear user-item interactions. {\bf (3)} The property of balancing the goals of exploration and exploitation alleviates the pressure of losing users caused by the full exploitation in existing meta-learning methods. Lastly, to verify its advantage over state-of-the-arts, we conduct extensive experiments and analysis on three benchmark datasets (MovieLens~\footnote{\url{https://grouplens.org/datasets/movielens/}}, EachMovie~\footnote{\url{https://www.librec.net/datasets.html}} and Netflix~\footnote{\url{https://www.kaggle.com/netflix-inc/netflix-prize-data}}). The experimental results demonstrate its significant advantage over state-of-the-art methods and the knowledge learned by NICF.

Our main contributions presented in this paper are as follows:
\begin{itemize}
  \item We formally propose to employ reinforcement learning to solve the cold-start and warm-start recommendation under the interactive collaborative filtering setting. 
  \item We propose to represent the exploration policy with multi-channel stacked self-attention neural networks and learn the policy network by maximizing users' satisfaction.
  \item We perform extensive experiments on three real-world benchmark datasets to demonstrate the effectiveness of our NICF approach and the knowledge learned by it.
\end{itemize}

\section{Preliminary}
In this section, we first formalize the interactive collaborative filtering on the perspective of the multi-armed bandit and then shortly recapitulate the widely used approaches and its limitations for interactive collaborative filtering. 

\subsection{A Multi-Armed Bandit Formulation}
In a typical recommender system, we have a set of $N$ users $U=\{1,\dots,N\}$ and a set of $M$ items $I = \{1,\dots,M\}$. The users' feedback for items can be represented by a $N\times M$ preference matrix $R$ where $r_{u,i}$ is the preference for item $i$ by user $u$. Here, $r_{u,i}$ can be either explicitly provided by the user in the form of rating, like/dislike, {\it etc,} or inferred from implicit interactions such as views, plays and purchases. In the explicit setting, $R$ typically contains graded relevance (\eg 1-5 ratings), while in the implicit setting $R$ is often binary. Without loss of generality, we consider the following process in discrete timesteps. At each timestep $t\in[0,1,2,\dots,T]$, the system delivers an item $i_{t}$ to the target user $u$, then the user will give feedback $r_{u,i_{t}}$, which represents the feedback collected by the system from user $u$ to the recommended item $i_t$ at timestep $t$. In other words, $r_{u,i_t}$ is the ``reward'' collected by the system from the target user. After receiving feedback, the system updates its model and decides which item to recommend next. Let's denote $s_t$ as the available information (the support set) the system has for the target user $s_t = \{i_1,r_{u,i_1},\dots,i_{t-1},r_{u,i_{t-1}}\}$ at timestep $t$.

Then, the item is selected according to a policy $\pi:s_t\rightarrow I$, which is defined as a function from the current support set to the selected item $i_t\sim\pi(s_t)$. In the interactive recommendation process, the total T-trial payoff of $\pi$ is defined as $\sum_{i=1}^T r_{t,i_t}$. For any user $u$, our goal is to design a policy $\pi$ so that the expected total payoff $G_\pi(T)$ is maximized,
\begin{eqnarray}\label{equ:target}
    G_\pi (T) = \mathbb{E}_{i_t\sim \pi(s_t)} \left[\sum_{t=1}^T r_{u,i_t}\right].
\end{eqnarray}
Similar, we can define the optimal expected $T$-trial payoff as $G^\ast (T) = \mathbb{E}\left[\sum_{t=1}^T r_{u,i^\ast_t}\right]$, where $i^\ast_t$ is the optimal recommendation with maximum expected reward at timestep $t$. Usually, in MAB, we would like to minimize the {\it regret} defined as $G^\ast (T) - G_\pi (T)$. However, in recommender system, it is more intuitive to directly maximize the cumulative reward $G_\pi (T)$, which is equivalent to minimize the {\it regret}.

\subsection{Multi-Armed Bandit Based Approaches}
\begin{figure}
\includegraphics[width=2.8in]{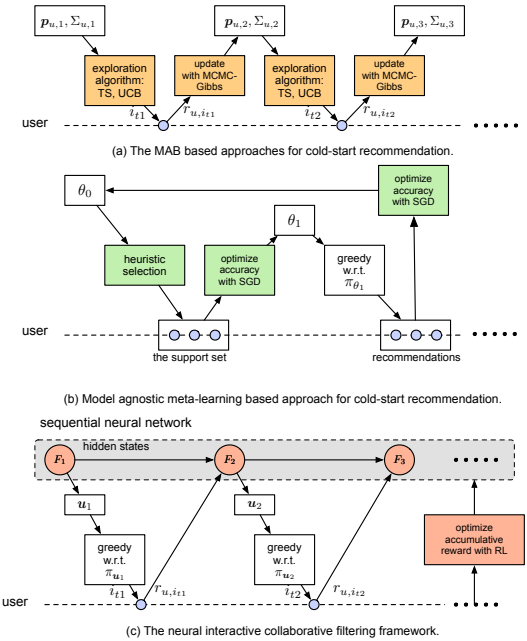}
    \caption{Difference between existing approaches and neural interactive collaborative filtering framework.}
    \label{fig:framework} 
\end{figure}

Currently, the exploration techniques in interactive collaborative filtering are mainly based on probabilistic matrix factorization (PMF)~\cite{mnih2008probabilistic}, which assumes the conditional probability distribution of rating follows a Gaussian distribution $Pr(r_{u,i}|\bm{p}_u^\top\bm{q}_i,\sigma^2) = \mathcal{N}(r_{u,i}|\bm{p}_u^\top\bm{q}_i,\sigma^2)$. Here, $\bm{p}_u$ and $\bm{q}_i$ are the user and item feature vectors with a zero mean Gaussian prior distribution and $\sigma$ is the prior variance. During the learning procedure, current approaches, as shown in Figure \ref{fig:framework} (a), iterate between two steps: \textbf{(1)} Obtaining the posterior distributions of the user and item feature vectors after the $(t-1)$-th interaction, denoting as $Pr(\bm{p}_u) = \mathcal{N}(\bm{p}_{u,t}|\bm{\mu}_{u,t},\Sigma_{u,t})$ and $Pr(\bm{q}_i) = \mathcal{N}(\bm{q}_{i,t}|\bm{\nu}_{i,t},\Psi_{i,t})$. The calculation of mean and variance terms $\{\bm{\mu}_{u,t}$, $\bm{\nu}_{i,t}$, $\Sigma_{u,t}$ and $\Psi_{i,t}\}$ can be obtained by following MCMC-Gibbs (refers to~\cite{zhao2013interactive}). \textbf{(2)} Heuristically select the item for the $t$-th recommendation with the aim of maximizing the cumulative reward. Specifically, there are mainly two strategies have been explored to select the items in interactive collaborative filtering:

\paragraph{Thompson Sampling~\cite{chapelle2011empirical}} At the timestep $t$ for user $u$, this method suggests the item with the maximum sampled values as $i_t = \underset{i}{\arg\max}$ $ \Tilde{\bm{p}}_{u,t}^\top\Tilde{\bm{q}}_{i,t}$, where $\Tilde{\bm{p}}_{u,t}\sim \mathcal{N}(\bm{\mu}_{u,t},\Sigma_{u,t})$ and $\Tilde{\bm{q}}_{i,t}\sim\mathcal{N}(\nu_{i,t},\Psi_{i,t})$ are sampled from the posterior distribution of user and item feature vectors~\cite{kaufmann2012thompson}.

\paragraph{Upper Confidence Bound} It based on the principle of optimism in the face of uncertainty, which is to choose the item plausibly liked by users. In~\cite{zhao2013interactive}, it designs a general solution Generalized Linear Model Bandit-Upper Confidence Bound (GLM-UCB), which combined UCB with PMF as
\begin{eqnarray*}
    i_t = \underset{i}{\arg \max }\left(\rho\left(\bm{\mu}_{u, t}^{\top} \bm{\nu}_{i,t}\right)+c \sqrt{\log t}\left\|\bm{\nu}_{i,t}\right\|_{2, \Sigma_{u, t}}\right).
\end{eqnarray*}
Here, $\rho$ is a sigmoid function defined as $\rho\left(x\right) = \frac{1}{1+\exp(-x)}$, $c$ is a constant with respect to $t$. $\left\|\bm{\nu}_{i,t}\right\|_{2, \Sigma_{u, t}}$ is 2-norm based on $\Sigma_{u, t}$ as $\left\|\bm{\nu}_{i,t}\right\|_{2, \Sigma_{u, t}} =\sqrt{\bm{\nu}_{i,t}^\top \Sigma_{u, t} \bm{\nu}_{i,t}}$, which measures the uncertainty of estimated rate $r_{u,i}$ at the $t$-th interaction. 

The above-discussed approaches show the possible limitation of MAB based methods: {\bf (1)} Owing to the difficulty of updating the posterior distribution for non-linear models, they are only applicable for linear user-item interaction models, which greatly limits its usage on effective neural networks based models~\cite{he2017neural,xue2017deep}. {\bf (2)} A lot of crucial hyper-parameters (\eg the variance term for prior distribution and the exploration hyper-parameter $c$) are introduced, which increases the difficulty of finding the optimal recommendations. {\bf (3)} The sophisticated approaches (Thompson Sampling and GLM-UCB) are potentially overly pessimistic since they are usually optimizing the recommendations in the worst case to achieve provably good regret bounds.  

\subsection{Meta-learning Based Approach}
Meta-learning based approaches aim to learn a learning procedure that can quickly capture users' interests after observed a small support set. As shown in Figure \ref{fig:framework} (b), we presented an example framework MELU~\cite{lee2019melu}, which adapted Model-Agnostic Meta-Learning (MAML)~\cite{finn2017model} for fastly model adaption on cold-start users. Specifically, assume the recommender agent is modeled with a neural network parameterized with $\theta$, MELU aims to learn an initialization $\theta_0$ that can identify users' interests after updating $\theta_1$ with small support set $D$. Formally, the $\theta_0$ is learned by minimizing a specific loss $\ell$ over the support set $D$ after updating to $\theta_1$ as
\begin{eqnarray*}
	\theta_1 &=& \theta_0 - \alpha \ell(\pi_{\theta_0},D) \\
	\theta_0 &\leftarrow & \theta_0 -\alpha \ell(\pi_{\theta_1},D), 
\end{eqnarray*} 
where $\pi_\theta$ is the recommendation policy parameterized by $\theta$. $\ell$ usually corresponds to an accuracy measure, such as {\it MSE} or {\it Cross entropy}. 

The meta-learning approach is appealing since it avoids the complexity of hand-designing the sophisticated exploration policies and enables us to take advantage of deep neural networks. However, how to select the support set without compromising users' experience has not been concerned in existing meta-learning approaches. It resulted in two problems: \textbf{(1)} It leads to the recommendation of highly irrelevant items and terrible user experience at the phase of constructing the support set. \textbf{(2)} These methods perform lousy when faced with users' tastes drifting or poor quality support set due to its full exploitation strategy and deficiency in actively exploring users' interests.

In the following, we address these limitations by employing a neural network based exploration policy, which directly learns to explore for interactive collaborative filtering.

\section{Neural Interactive Collaborative Filtering}
We first present the general neural interactive collaborative filtering framework, elaborating how to formulate the exploration in cold-start and warm-start recommendation as a meta RL task, a bandit problem within an MDP. To explore DNNs for modeling user-item interactions, we then propose an instantiation of NICF, using stacking self-attention neural networks to represent the recommendation policy under interactive collaborative filtering. Lastly, we present an efficient policy learning method for interactive collaborative filtering.

\subsection{General Framework}
Rather than hand-designing exploration strategies for cold-start or warm-start users, we take a different approach in this work and aim to learn a neural network based exploration strategy whereby the recommender agent can capture users' interests rapidly for different users and hence maximize the cumulative users' engagement in the system, \ie we would like to learn a general procedure (a sequential neural network) that takes as input a set of items from any user's history and produces a scoring function that can be applied to new test items and balance the goals between learning the user profile and making accurate recommendations (as shown in Figure \ref{fig:framework}(c)). 

In this formulation, we notice that the interactive collaborative filtering is equivalent to a meta-learning problem where the objective is to learn a learning algorithm that can take as the input of the user's history $s_t$ and will output a model (policy function) that can be applied to new items. From the perspective of meta-learning, the neural network based policy function is a low-level system, which learns quickly and is primarily responsible for exploring users' interests, and we want to optimize the low-level system with a slower higher-level system that works across users to tune and improve the lower-level system~\cite{duan2016rl}. Specifically, for every user $u$, the agent executes a sequential neural network based policy $\pi_{\theta}(s_t)$, which constantly updates its recommendation policy based on the recommending items and users' feedback. The slower higher-level system optimizes the weights of the sequential neural network in an end-to-end way to maximize the cumulative reward $G_\pi(T)$, which can be viewed as a reinforcement learning problem and optimized with RL algorithm. 

From the perspective of RL, applying RL to solve cold-start and warm-start recommendation is also meaningful since the users' preferences gathered by exploration recommendations can trigger much more satisfied recommendations, which can be viewed as the delayed reward for the recommendations and RL is born to maximize the sum of delayed and immediate reward in a global view. Therefore, applying RL directly achieves the goal of balancing between exploration and exploitation for interactive collaborative filtering. In details, as a RL problem,  $\langle S , A , P, R, \gamma \rangle$ in the MDP are defined as: \textbf{(1)} State $S$ is a set of states, which is set as the support set $s_t \in S$. \textbf{(2)}  Action set $A$ is equivalent to item set $I$ in recommendation. \textbf{(3)} Transition $P$ is the transition function with $Pr \left( s_{t+1} | s_t , i_t \right)$ being the probability of seeing state $s_{t+1}$ after taking action $i_t$ at $s_t$. In our case, the uncertainty comes from user's rating $r_{u,i_t}$ \emph{w.r.t.} $i_t$ and $s_t$. \textbf{(4)} Reward $R$ is set based on users' feedback, \ie the user's rating.

\subsection{Self-Attentive Neural Policy}
\begin{figure}
\includegraphics[width=3.3in]{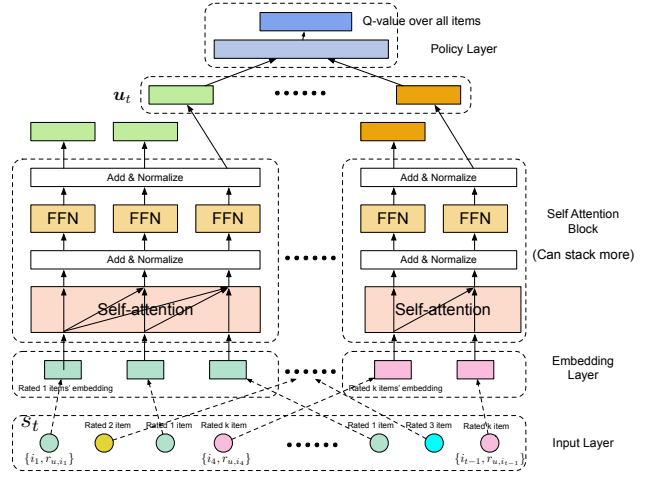}
    \caption{The neural architecture for the recommender agent.}
    \label{fig:network} 
\end{figure}

In this work, the exploration policy is parameterized with multi-channel stacked self-attention neural networks, which separately capture the information of versatile user behaviors since different rewarding recommendations for a specific user are usually extremely imbalanced (\eg liking items usually are much fewer than disliking items)~\cite{zou2019reinforcement,zhao2018recommendations,kang2018self}. In Figure \ref{fig:network}, we presented the neural architecture for exploration policy, which consists of an embedding layer, self-attentive blocks, and a policy layer.

\paragraph{Embedding Layer} 
Given $s_t=\{i_1,r_{u,i_1},\dots,i_{t-1},r_{u,i_{t-1}}\}$, the entire set of $\{i_t\}$ are converted into item embedding vectors $\{\bm{i}_t\}$ of dimension $d$ by embedding each $i_t$ in a continuous space, which, in the simplest case, is an embedding matrix $A\in \mathbb{R}^{I\times d}$. 

\paragraph{Self-Attention Layer} 
To better represent the observation $s_t$, as shown in Figure \ref{fig:network}, we separately process different rated items by employing multi-channel stacked self-attentive neural networks. Denote the items rated with score $z$ as an embedding matrix as $E^z_{t} = \left[\cdots,\bm{i}_m,\cdots\right]^\top$ ($\forall r_{u,i_m} = z, m<t$). The self-attention operation takes the embedding $E^z_{t}$ as input, converts it to three matrices through linear projects, and feeds them into an attention layer
\begin{eqnarray*}
	S^{z}_t = \text{SA} (E^z_{t}) = \texttt{Attention} (E^z_{t} W^{z,c}, E^z_{t} W^{z,k}, E^z_{t} W^{z,v}),
\end{eqnarray*}
where $W^{z,c}, W^{z,k}, W^{z,v} \in \mathbb{R}^{d\times d}$ are the projection matrices. These projections make the model more flexible. $\texttt{Attention}$ function is the scaled dot-product attention
\begin{eqnarray*}
	\texttt{Attention}(C,K,V) = \text{softmax}\left(\frac{CK^\top}{\sqrt{h}}\right)V,
\end{eqnarray*}
where $C$ represents the queries, $K$ the keys and $V$ the values (each row represents an item). The scale factor $\sqrt{h}$ is to avoid overly large values of the inner product, especially when dimensionality is high. Due to sequential nature of the recommendations, the attention layer should only consider the first $t-1$ items when formulating the $t$-th policy. Therefore, we modify the attention by forbidding all links between $
C_i$ and $K_j$ ($j>i$).

\paragraph{Point-Wise Feed-Forward Layer} To endow the model with nonlinearity and to consider interactions between different latent dimensions, we apply a point-wise two-layer feed-forward network to ${S^{z}_{t}}_m$ (the $m$-th row of the self-attention layer $S^{z}_t$) as
\begin{eqnarray*}
	{F^{z}_{t}}_m = \text{FFN}({S^{z}_{t}}_m) = \text{ReLU} ({S^{z}_{t}}_m W^{(1)} + \bm{b}^{(1)}) W^{(2)} + \bm{b}^{(2)},
\end{eqnarray*}
where $\text{ReLU}(x)=\max(0,x)$ is the rectified linear unit. $W^{(1)}$ and $W^{(2)}$ are $d\times d$ matrics. $\bm{b}^{(1)}$ and $\bm{b}^{(2)}$ are $d$-dimensional vectors. 

\paragraph{Stacking Self-Attention Block} The self-attention layer and point-wise feed-forward layer, which formulates a self-attention block and can be stacked to learn more complex item transitions. Specifically, $b$-th ($b>1$) block is defined as:
\begin{eqnarray*}
	S^{z,b}_t &=& \text{SA}(F^{z,b-1}_{t}),\\
	{F^{z,b}_{t}}_{m} &=&  \text{FFN}({S^{z,b}_{t}}_m)
\end{eqnarray*}
and the $1$-st block is defined as $S^{z,1}_t = S^{z}_t$ and $F^{z,1}_t = F^{z}_t$.

\paragraph{Policy Layer}
After $b$ self-attention blocks that adaptively and hierarchically extract information of previously rated items, we predict the next item score based on $\left\{F^{z,b}_{t}\right\}^{R_{\text{max}}}_{z=1}$, where $R_{\text{max}}$ is the maximal reward. Denote the predicted cumulative reward of recommending items as $\bm{Q}_{\theta}(s_t,\cdot) = [Q_{\theta}(s_t,i_1),\cdots,Q_{\theta}(s_t,i_N)]^\top$, the policy layer is processed by two feed-forward layers as,
\begin{eqnarray*}
\bm{u}_t  &=& \text{concat}\left[{F^{1,b}_{t}}^\top,{F^{2,b}_{t}}^\top,\dots,{F^{R_{\text{max}},b}_{t}}^\top\right]^\top \\
\bm{Q}_{\theta}(s_t,\cdot) &=&\text{ReLU} ( \bm{u}_t W^{(1)} + \bm{b}^{(1)}) W^{(2)} + \bm{b}^{(2)},
\end{eqnarray*}
where $W^{(1)}\in \mathbb{R}^{R_{\text{max}}d\times d}$, $W^{(2)}\in \mathbb{R}^{d\times |I|}$ are weight matrices and $\bm{b}^{(1)}\in \mathbb{R}^d$ and $\bm{b}^{(2)}\in \mathbb{R}^{|I|}$ are the bias terms. With the estimated $\bm{Q}_{\theta}(s_t,\cdot)$, the recommendation is generated by selecting the item with maximal Q-value as $\pi_{\theta}(s_t) = \underset{i}{\arg\max}\,Q_{\theta}(s_t,i)$. 

\subsection{Policy Learning}
\paragraph{Q-Learning}
We use Q-Learning~\cite{mnih2013playing} to learn the weights $\theta$ for the exploration policy. In the $t$-th trial, the recommender agent observes the support set $s_t$, and chooses the item $i_t$ with an $\epsilon$-greedy policy \emph{w.r.t.} the approximated value function $\bm{Q}_{\theta}(s_t,\cdot)$ (\ie with probability $1-\epsilon$ selecting the max Q-value action, with probability $\epsilon$ randomly choosing an action). The agent then receives the response $r_{u,i_t}$ from the user and updates the observed set to $s_{t+1}$. Finally, we store the experience $(s_t,a_t,r_{u,i_t},s_{t+1})$ in a large replay buffer $\mathcal{M}$ from which samples are taken in mini-batch training.

We improve the value function $Q_{\theta}(s_t,i_t)$ by adjusting $\theta$ to minimize the mean-square loss function, defined as follows:
\begin{eqnarray}\label{equ:dqn}
\ell(\theta) &=& \mathbb{E}_{(s_t,i_t,r_{u,i_t},s_{t+1})\sim \mathcal{M}}\left[(y_t-Q_{\theta}(s_t,i_t))^2\right]\\ \nonumber
    y_t &=& r_{u,i_t}+ \gamma \max_{i_{t+1}\in \mathcal{I}} Q_{\theta}(s_{t+1},i_{t+1}),
\end{eqnarray}
where $y_t$ is the target value based on the optimal {\it Bellman Equation}~\cite{sutton2018reinforcement}. 
By differentiating the loss function \emph{w.r.t.} $\theta$, we arrive at the following gradient:
\begin{eqnarray}\label{equ:qlearning}
\nonumber \nabla _ { \theta } \ell \left( \theta  \right) &=& \mathbb { E } _ { \left(s_t,i_t,r_{u,i_t},s_{t+1}\right) \sim \mathcal {M} } \left[(y_t - Q_\theta \left( s_t , i_t\right) \right)\nabla _ { \theta  } Q_\theta \left( s_t , i_t)\right].
\end{eqnarray}

\paragraph{Efficient Learning}
Usually, training a RL agent is much more challenging than supervised learning problems~\cite{sutton2018reinforcement}. Additionally, in recommender systems, the large-scale action space and state space have greatly increased the difficulty of training a reinforcement learning-based recommender agent~\cite{zou2020dynaq,chen2019top}. To reduce the difficulty, we adapt a constantly increased $\gamma$ during the training as $\gamma_{e} = \frac{1}{1+(E-e)^{\eta}}$, where $e$ is the $e$-th epoch, $E$ is the total number of epoch, and $\eta$ is a hyper-parameter (we set $\eta=0.2$ in the experiments). Since the larger $\gamma$ means planning in longer future horizons for RL, the increasing $\{\gamma_e\}$ can be treated as an increasingly difficult curriculum~\cite{bengio2009curriculum}, which gradually guides the learning agent from 1-horizon (greedy solution), 2-horizon, $\dots$, to overall optimal solutions. Therefore, it is much more efficient than finding the optimal recommender policy from scratch. 
\section{Experiments}
In this section, we conduct extensive experiments on three benchmark datasets to evaluate the effectiveness of NICF. We mainly focus on answering the following research questions:

\textbf{RQ1}: How can NICF outperform existing interactive collaborative filtering algorithms for the cold-start users?

\textbf{RQ2}: Can the NICF be applied to warm-start users with drifting taste, \ie those whose interests change over time?

\textbf{RQ3}: What's the influence of various components in NICF?

\textbf{RQ4}: What kind of knowledge learned by NICF for cold-start recommendations? 

In what follows, we will first introduce our experimental settings, followed by answering the above four research questions.

\subsection{Experimental Settings}
\subsubsection{Datasets}
We experiment with three real-world benchmark datasets: MovieLens 1M\footnotemark[4], EachMovie\footnotemark[5], and Netflix\footnotemark[6]. Table~\ref{tab:data_info} lists the statistics of the three datasets.

Due to the interactive nature of the recommender system, an online experiment with true interactions from real users would be ideal, but it is not always possible~\cite{li2010contextual,zhao2013interactive}. Following the setting of interactive collaborative filtering~\cite{zhao2013interactive,he2017neural}, we assume that the ratings recorded in the datasets are users' instinctive actions, not biased by the recommendations provided by the system. In this way, the records can be treated as unbiased to represent the feedback in an interactive setting. Additionally, we assume that the rating is no less than 4 is the satisfied recommendation, otherwise dissatisfied. These assumptions define a simulation environment for training and evaluating our proposed algorithm and the learning agent is expected to keep track of users' interests and recommend successful items throughout a long time.

\subsubsection{Compared Methods}
We compare our model with state-of-the-art methods from different types of recommendation approaches, including:
\begin{itemize}
\item \textbf{Random}: The random policy is executed in every recommendation, which is a baseline used to estimate the worst performance that should be obtained.

\item \textbf{Pop}: It ranks the items according to their popularity measured by the number of being rated. This is a widely used simple baseline. Although it is not personalized, it is surprisingly competitive in evaluation, as users tend to consume popular items. 

\item \textbf{MF}~\cite{koren2009matrix}: It suggests recommendations based on the ratings of other users who have similar ratings as the target user. For cold-start recommendation, we always greedy \emph{w.r.t.} the estimated scores and update users' latent factor after every interaction. 

\item \textbf{MLP}: Multi-layer perceptron has been a common practice for non-linear collaborative filtering~\cite{he2017neural,xue2017deep} due to its superiority. We deploy a MLP based recommender agent using the architecture mentioned in~\cite{he2017neural}.

\item \textbf{BPR}~\cite{rendle2009bpr}: It optimizes the MF model with a pairwise ranking loss, which is a state-of-the-art model for item recommendation.

\item \textbf{ICF}~\cite{zhao2013interactive}: Interactive collaborative filtering combined the probabilistic matrix factorization~\cite{mnih2008probabilistic} with different exploration techniques for recommender system, including GLM-UCB (generalized LinUCB~\cite{li2010contextual}), TS~\cite{chapelle2011empirical} and $\epsilon$-Greedy~\cite{sutton2018reinforcement}, which are strong baselines for handling exploration/exploitation dilemma in recommender system. 

\item \textbf{MeLU}~\cite{lee2019melu}: MeLU is a state-of-the-art method, which adapted MAML~\cite{finn2017model} for solving the cold start problem by treating it as a few-shot task. 

\item \textbf{NICF}: Our proposed approach for learning to explore in cold-start or warm-start recommendation.
\end{itemize}

\begin{table}[!t]
\centering
\tabcolsep 0.03in
\caption{Summary Statistics of Datasets.}
\begin{tabular}{l|c|c|c}
\ChangeRT{1pt}
Dataset & MovieLens (1M) & EachMovie & Netflix\\ \hline\hline
\# Users & 6,040 & 1,623 & 480,189 \\
\# Items & 3,706 & 61,265 & 17,770 \\
\# Interactions & 1,000,209 & 2,811,718 & 100,480,507 \\
\# Interactions Per User & 165.60 & 1732.42 & 209.25 \\
\# Interactions Per Item & 269.89 & 45.89 & 5654.50 \\
\ChangeRT{1pt}
\end{tabular}
\label{tab:data_info}
\end{table}

\subsubsection{Evaluation Metrics} Given a cold-start or warm-start user, a well-defined exploration strategy should recommend the items to deliver the maximal amount of information useful for estimating users' preferences. Previously, this kind of exploration is achieved by improving the diversity of recommendations~\cite{cheng2017learning,zou2019diversity}. Hence, to study the learned exploration strategy, we evaluate the model on both the accuracy and diversity of generated recommendations. Given the ordered list of items, we adopt three widely used metrics in recommender system:
\begin{itemize}
    \item \textbf{Cumulative Precision$@T$}. A straightforward measure is the number of positive interactions collected during the total $T$ interactions,
    \begin{eqnarray}
         \text{precision}@T = \frac{1}{\text{\#  users}}\sum_{\text{users}}\sum_{t=1}^T b_t.
    \end{eqnarray}
    For both datasets, we define $b_t = 1$ if $r_{u,i_t}>=4$, and $0$ otherwise.
    \item \textbf{Cumulative Recall$@T$}. We can also check for the recall during $T$ timesteps of the interactions,
    \begin{eqnarray}
        \text{recall}@T = \frac{1}{\text{\# users}}\sum_{\text{users}}\sum_{t=1}^{T}\frac{b_t}{\text{\# \text{satisfied items}}}.
    \end{eqnarray}
    \item \textbf{Cumulative $\alpha\text{-}NDCG@T$}. $\alpha\text{-}NDCG@T$ generalize $NDCG@T$ to diversity of the recommendation list, which formulated as
    \begin{eqnarray}
        \alpha\text{-}NDCG@T &=& \frac{1}{\mathcal{Z}} \sum_{t=1}^T  \frac{G@t}{\log (1+t)}.
    \end{eqnarray}
Here, $G@t=\sum_{\forall i \in C}(1-\alpha)^{c_{i, t}-1}$ with $c_{i,t}$ as the number of times that topic $i$ has appeared in the ranking of the recommendation list up to (and including) the $t$-th position. Here, the topic is the property of items or users. $\mathcal{Z}$ is the normalization factor.
\end{itemize}

\subsubsection{Parameter Setting} These datasets are split into three user-disjoint sets: 85\% users' data as the training set and their ratings are used to learn the parameters for the models, 5\% users' data used for tuning hyper-parameters, including the learning rate, hidden units, and early stop. The last 10\% of users go through the interactive recommendation process during 40 time-steps which are used to evaluate the effectiveness of different methods. For all methods except Random and Pop, grid search is applied to find the optimal settings. These include latent dimensions $d$ from $\{10,20,30,40,50\}$, and the learning rate from $\{1,0.1,0.01,0.001,0.0001\}$. We report the result of each method with its optimal hyper-parameter settings on the validation data. We implement our proposed methods with Tensorflow and the code is available at \url{https://github.com/zoulixin93/NICF}. The optimizer is the {\it Adam} optimizer~\cite{kingma2014adam}. We stack two self-attentive blocks in the default setting. The capacity of the replay buffer for Q-learning is set to 10000 in experiments. The exploration factor $\epsilon$ decays from 1 to 0 during the training of the neural network. 


{\centering
\begin{table*}[!hpt]
\centering
\tabcolsep 0.04in
\caption{Cold-start recommendation performance of different models on MovieLens (1M), EachMovie and Netflix Dataset.}
\label{tab:cold}
\begin{tabular}{c| c c c c| c c c c| c c c c} 
\ChangeRT{0.8pt}
Dataset &\multicolumn{4}{c|}{MovieLens (1M)} & \multicolumn{4}{c|}{EachMovie}  & \multicolumn{4}{c}{Netflix}\\\hline\hline
Measure & \multicolumn{4}{c|}{Cumulative Precision} & \multicolumn{4}{c|}{Cumulative Precision} & \multicolumn{4}{c}{Cumulative Precision} \\\hline
T & 5 & 10 & 20 & 40 & 5 & 10 & 20 & 40 & 5 & 10 & 20 & 40  \\ \hline\hline
Random            & 0.2150 & 0.4400 & 0.8983 & 1.8100  & 0.2454 & 0.4663 & 0.7730  & 1.4233  & 0.0600 & 0.1267 & 0.2683 & 0.5000 \\
Pop               & 2.4933 & 4.6383 & 8.6267 & 15.6100 & 4.0123 & 6.3497 & 10.2699 & 17.2699 & 2.1283 & 4.0217 & 7.3183 & 13.4067 \\
MF                & 2.6947 & 4.9684 & 9.1579 & 16.0947 & 4.0534 & 6.3167 & 10.3582 & 17.6167 & 2.4667 & 4.5500 & 8.3333 & 14.9500 \\
BPR               & 2.9579 & 5.4842 & 9.7895 & 16.6526 & 4.0534 & 6.4552 & 10.4598 & 17.9333 & 2.2833 & 4.5512 & 8.4532 & 15.3667 \\
MLP               & 2.7158 & 5.3895 & 9.8105 & 16.9158 & 4.1041 & 6.9384 & 11.2740 & 18.8425 & 2.5491 & 4.8966 & 8.7241 & 15.9077 \\

\hline\hline
$\epsilon$-greedy & 2.9714 & 5.6286 & 10.4286 & 17.1429 & 4.1126 & 6.9790 & 11.3846 & 19.0420 & 2.6875 & 5.1312 & 9.1250 & 16.0438 \\
TS                & 3.0968 & 5.8713 & 11.0323 & 18.3548 & 4.1596 & 7.6422 & 13.0020 & 22.7431 & 2.7841 & 5.3864 & 9.6818 & 17.2841 \\
GLM-UCB            & 3.2917 & 6.2083 & 11.5833 & 19.0932 & 4.1761 & 7.8586 & 13.5556 & 23.9293 & 2.8739 & 5.4752 & 9.9375 & 17.9125 \\
MELU              & 3.3636 & 6.3182 & 11.9545 & 19.7273 & 4.1316 & 7.8421 & 13.3816 & 23.9605 & 2.8298 & 5.4711 & 9.8541 & 17.3951 \\
NICF              & ${3.5556}^\ast$ & ${6.7778}^\ast$ & ${12.9444}^\ast$ & ${21.5875}^\ast$ & ${4.2270}^\ast$ & ${7.8957}^\ast$ & ${14.5215}^\ast$ & ${25.0613}^\ast$ & ${2.9641}^\ast$ & ${5.7647}^\ast$ & ${10.4542}^\ast$ & ${18.5523}^\ast$ \\
\hline
\hline
Measure & \multicolumn{4}{c|}{Cumulative Recall} & \multicolumn{4}{c|}{Cumulative Recall} & \multicolumn{4}{c}{Cumulative Recall} \\\hline
T & 5 & 10 & 20 & 40 & 5 & 10 & 20 & 40 & 5 & 10 & 20 & 40  \\ \hline\hline
Random            & 0.0011 & 0.0027 & 0.0051 & 0.0106 & 0.0001  & 0.0001 & 0.0003 & 0.0004 & 0.0003 & 0.0007 & 0.0014 & 0.0025 \\
Pop               & 0.0268 & 0.0443 & 0.0797 & 0.1375 & 0.0445 & 0.0541 & 0.0906 & 0.1295 & 0.0215 & 0.0390 & 0.0672 & 0.1152 \\
MF                & 0.0300 & 0.0497 & 0.0823 & 0.1443 & 0.0477 & 0.0536 & 0.0908 & 0.1301 & 0.0247 & 0.0454 & 0.0749 & 0.1198 \\
BPR               & 0.0353 & 0.0534 & 0.0926 & 0.1483 & 0.0477 & 0.0592 & 0.0911 & 0.1321 & 0.0233 & 0.0459 & 0.0758 & 0.1201 \\
MLP               & 0.0305 & 0.0526 & 0.0961 & 0.1490 & 0.0485 & 0.0709 & 0.1010 & 0.1360 & 0.0258 & 0.0472 & 0.0775 & 0.1220 \\
\hline\hline
$\epsilon$-greedy & 0.0358 & 0.0572 & 0.1083 & 0.1522 & 0.0490 & 0.0712 & 0.1062 & 0.1392 & 0.0264 & 0.0482 & 0.0788 & 0.1241  \\
TS                & 0.0371 & 0.0601 & 0.1138 & 0.1693 & 0.0493 & 0.0798 & 0.1102 & 0.1452 & 0.0270 & 0.0508 & 0.0817 & 0.1280\\
GLM-UCB            & 0.0382 & 0.0614 & 0.1147 & 0.1853 & 0.0507 & 0.0817 & 0.1120 & 0.1488 & 0.0281 & 0.0524 & 0.0862 & 0.1332 \\
MELU              & 0.0389 & 0.0639 & 0.1173 & 0.1971 & 0.0501 & 0.0810 & 0.1113 & 0.1505 & 0.0274 & 0.0519 & 0.0855 & 0.1292 \\
NICF              & ${0.0409}^\ast$ & ${0.0652}^\ast$ & ${0.1202}^\ast$ & ${0.2145}^\ast$ & ${0.0511}^\ast$ & ${0.0821}^\ast$ & ${0.1195}^\ast$ & ${0.1523}^\ast$ & ${0.0284}^\ast$ & ${0.0535}^\ast$ & ${0.0901}^\ast$ & ${0.1374}^\ast$ \\
\ChangeRT{0.8pt} 
\multicolumn{13}{c}{$``\ast"$ indicates the statistically significant improvements (\ie two-sided $t$-test with  $p<0.05$) over the best baseline.} 
\end{tabular}
\end{table*}
}

\subsection{Performance comparison on cold-start cases (RQ1)}
Table \ref{tab:cold} reports the performance of accumulative precision and recall throughout 40 trial recommendations for cold-start cases. The results are quite consistent with our intuition. We have the following observations: 

\textbf{(1)} Our method NICF outperforms other baselines on three benchmark datasets. We can see that NICF achieves the best performance on the precision and recall over three benchmark datasets, significantly outperforming the state-of-the-art methods by a large margin (on average, the relative improvement on cumulative precision@40 over the best baseline are 9.43\%, 4.59\% and 6.65\% for three benchmark datasets, respectively). It means that for cold-start recommendation, our proposed method can quickly capture users' interests, and adapt its strategy to cater to new users.

\textbf{(2)} The GLM-UCB and TS algorithms generally work better than the greedy methods MF, BRP, MLP, and heuristic search method $\epsilon$-greedy. In most cases, TS and GLM-UCB also exceed other baseline algorithms on EachMovie and Netflix datasets (according to the cumulative precision and recall). It means that the exploration by considering the uncertainties of the user and items according to their probability distributions is more promising than random explorations. Nevertheless, TS and GLM-UCB fail to outperform our proposed NICF algorithms.

\textbf{(3)} Overall, the meta-learning method, MELU, consistently outperforms the traditional baselines on average as shown in Table \ref{tab:cold}, and is much better than all other baselines on MovieLen (1M), which indicates that meta-learning method helps improve the recommendation accuracy on cold-start recommendation.

\subsection{Performance comparison on warm-start cases with taste drift (RQ2)}
Through this experiment, we aim to answer the question of whether the algorithms are also applicable to warm-start users to follow up their interests throughout the interactions, especially when their tastes are changing over time. To do this, we first divide the rating records of the users (whose ratings are more than 80) into two periods (set 1 and set 2). For the selected user, the set 1 (20 items) is used as the historical interactions for the user and set 2 as the simulation for his/her taste drift. Then, we employ the genre information of the items as an indication of the user interest~\cite{zhao2013interactive}. That is, we calculate the cosine similarity between the genre vectors of the two periods. We choose the users with the smallest cosine similarity as an indication that they have significant taste drifting across the two periods. Since the genre information of EachMovie is not available, we only conduct experiments on MovieLens (1M) and Netflix datasets (the genre of Netflix dataset is crawled by using IMDBpy\footnote{\url{https://github.com/alberanid/imdbpy}}). Specifically, we respectively selected 4,600 users and 96,037 users from MovieLens (1M) and Netflix datasets to train and evaluate on warm-start recommendations.

Table \ref{tab:warm} reports the performance of accumulative precision and recall throughout 40 trial recommendations for warm-start users with drifting interests. In Table \ref{tab:warm}, it can be seen that our proposed methods outperform the baselines for both datasets. When compared with the best baseline, the improvement is up to 7.92\% on MovieLens (1M) dataset, and 6.43\% on the Netflix dataset, which means that for warm-start users, our proposed method can keep track on users' drifting taste and adapt its strategy to cater to users.

{\small
\begin{table}[!hpt]
\centering
\tabcolsep 0.001in
\caption{Warm-start recommendation performance of different models on MovieLens (1M) and Netflix Dataset.}
\label{tab:warm}
\begin{tabular}{c| c c c c | c c c c} 
\ChangeRT{0.8pt}
Dataset &\multicolumn{4}{c|}{MovieLens (1M)} &\multicolumn{4}{c}{Netflix} \\\hline\hline
Measure & \multicolumn{4}{c|}{Cumulative Precision} & \multicolumn{4}{c}{Cumulative Precision} \\\hline
T & 5 & 10 & 20 & 40 & 5 & 10 & 20 & 40  \\ \hline\hline
Random            & 0.2779 & 0.5284 & 1.0232 & 1.9305  & 0.0724 & 0.1281 & 0.2953 & 0.5877  \\
Pop               & 1.8589 & 3.6105 & 6.8926 & 12.4758 & 1.8162 & 3.6128 & 6.7883 & 13.3120   \\
MF                & 2.2527 & 4.4416 & 8.1444 & 14.5966 & 1.9466 & 3.8221 & 7.3381 & 14.1708  \\
BPR               & 2.3850 & 4.7084 & 8.6651 & 15.5513 & 2.1325 & 4.0602 & 7.4819 & 14.3373  \\
MLP               & 2.4654 & 4.8640 & 8.9881 & 16.1504 & 2.0784 & 4.0941 & 7.9098 & 15.3098  \\
\hline\hline
$\epsilon$-greedy & 2.5198 & 4.9851 & 9.2302 & 16.6015  & 2.2817 & 4.3803 & 8.2676 & 15.8310 \\
TS                & 2.6570 & 5.2190 & 9.6623 & 17.3668  & 2.2788 & 4.4779 & 8.7389 & 16.8805 \\
GLM-UCB           & 2.8237 & 5.5491 & 10.2688 & 18.4509 & 2.3505 & 4.6121 & 9.0374 & 17.4907 \\
MELU              & 2.8237 & 5.5051 & 10.2120 & 18.3220 & 2.6230$^\ast$ & 4.9672 & 9.4918 & 18.1475\\
NICF              & ${3.0097}^\ast$ & ${5.9385}^\ast$ & ${10.9935}^\ast$ & ${19.7735}^\ast$ & $2.6077$ & ${5.1215}^\ast$ & ${10.0829}^\ast$ & ${19.3149}^\ast$ \\
\hline
\hline
Measure & \multicolumn{4}{c}{Cumulative Recall} & \multicolumn{4}{c}{Cumulative Recall} \\\hline
T & 5 & 10 & 20 & 40 & 5 & 10 & 20 & 40  \\ \hline\hline
Random            & 0.0017 & 0.0031 & 0.0057 & 0.0105 & 0.0005 & 0.0008 & 0.0014 & 0.0028 \\
Pop               & 0.0144 & 0.0273 & 0.0506 & 0.0892 & 0.0107 & 0.0202 & 0.0386 & 0.0809 \\
MF                & 0.0189 & 0.0370 & 0.0640 & 0.1110 & 0.0110 & 0.0211 & 0.0391 & 0.0814 \\
BPR               & 0.0193 & 0.0379 & 0.0663 & 0.1150 & 0.0113 & 0.0215 & 0.0397 & 0.0819 \\
MLP               & 0.0201 & 0.0389 & 0.0672 & 0.1190 & 0.0111 & 0.0217 & 0.0402 & 0.0821 \\
\hline\hline
$\epsilon$-greedy & 0.0209 & 0.0396 & 0.0679 & 0.1200 & 0.0111 & 0.0206 & 0.0361 & 0.0671 \\
TS                & 0.0213 & 0.0403 & 0.0684 & 0.1208 & 0.0110 & 0.0210 & 0.0371 & 0.0716 \\
GLM-UCB            & 0.0219 & 0.0410 & 0.0692 & 0.1213 & 0.0116 & 0.0218 & 0.0378 & 0.0726 \\
MELU              & 0.0219 & 0.0412 & 0.0690 & 0.1210 & 0.0130$^\ast$ & 0.0223 & 0.0381 & 0.0740 \\
NICF              & ${0.0224}^\ast$ & ${0.0420}^\ast$ & ${0.0709}^\ast$ & ${0.1300}^\ast$ & 0.0128 & ${0.0228}^\ast$ & ${0.0390}^\ast$ & ${0.0752}^\ast$ \\
\ChangeRT{0.8pt}
\multicolumn{9}{c}{$``\ast"$ indicates the statistically significant improvements}\\
\multicolumn{9}{c}{(\ie two-sided $t$-test with  $p<0.05$) over the best baseline.}\\
\end{tabular}
\end{table}}

{
\begin{table}[!hpt]
\newcommand{\tabincell}[2]{\begin{tabular}{@{}#1@{}}#2\end{tabular}}
\centering
\tabcolsep 0.08in
\caption{Ablation analysis (Cumulative Precision$@40$) on three benchmark datasets. Performance better than default version is boldfaced. '$\downarrow$' indicates a severe performance drop (more than $10\%$).}
\label{tab:deep}
\begin{tabular}{c| c c c} 
\ChangeRT{0.8pt}
Architecture & MovieLens(1M) & EachMovie & Netflix\\\hline\hline
Default       & 21.5875 & 25.0613 & 18.5523 \\
LSTM          & 20.7895 & 23.2881 & 17.8185 \\
$\gamma = 0$  & 19.7273$\downarrow$ & 23.1656 & 17.1429 \\
0 Block (b=0) & 16.7368$\downarrow$ & 17.0276$\downarrow$ & 14.1250$\downarrow$ \\
1 Block (b=1) & 20.9818 & 24.9333 & 18.0429 \\
3 Block (b=3) & 21.4544 & \textbf{25.1063} & \textbf{18.6074} \\
Multi-Head    & 21.4167 & 24.2207 & 18.1502 \\
\ChangeRT{0.8pt}
\end{tabular}
\end{table}}

\subsection{Ablation Study (RQ3)}
Since there are many components in our framework, we analyze their impacts via an ablation study. Table \ref{tab:deep} shows the performance of our default method and its 4 variants on three datasets (with d = 30). We introduce the variants and analyze their effect respectively:

(1) LSTM: Replacing the self-attention blocks with LSTM cell, which is used to verify the effectiveness of self-attention on interactive collaborative filtering. Specifically, we adopt a two-layer LSTM with the hidden dimension of 30. The results imply that applying stacked self-attention blocks is beneficial for interactive collaborative filtering.

(2) $\gamma=0$: $\gamma=0$ means learning without using RL, \ie training a multi-channel stacked self-attention recommendation policy without consideration about the delayed reward, \ie the model delivers items in full exploitation way without consideration of exploration. Not surprisingly, results are much worse than the default setting.

(3) Number of blocks: Not surprisingly, results are inferior with zero blocks, since the model would only depend on the last item. The variant with one block performs reasonably well and three blocks performance a little better than two blocks, meaning that the hierarchical self-attention structure is helpful to learn more complex item transitions.

(4) Multi-head: The authors of Transformer~\cite{vaswani2017attention} found that it is useful to use 'multi-head' attention. However, performance with two heads is consistently and slightly worse than single-head attention in our case. This might owe to the small $d$ in our problem (d = 512 in Transformer), which is not suitable for decomposition into smaller subspaces.

\subsection{Analysis on Diversity (RQ4)}\label{sec:case}
\begin{figure}
\includegraphics[width=2.8in]{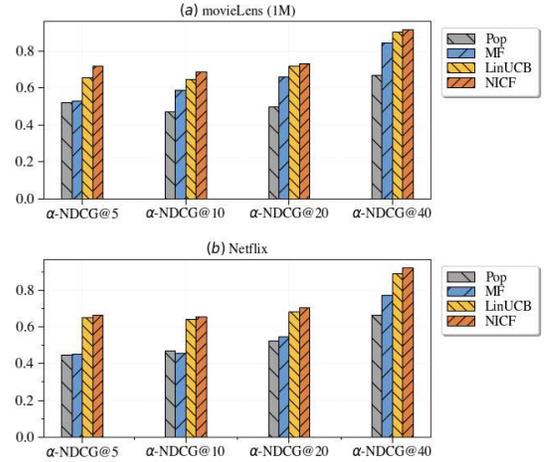}
    \caption{The recommendation diversity on cold-start phase.}
    \label{fig:diversity} 
\end{figure}

\paragraph{Diversity and accuracy}
Some existing works~\cite{cheng2017learning} explore users' interests by improving the recommendation diversity. It is an indirect method to keep exploration, and the assumption has not been verified. Intuitively, the diverse recommendation brings more information about users' interests or item attributes. Here, we conduct experiments to see whether NICF, which directly learn to explore, can improve the recommendation diversity. Since the genre information is only available on MovieLens (1M) and Netflix, we mainly analyze the recommendation diversity on these two datasets. In Figure \ref{fig:diversity}, the accumulative $\alpha$-NDCG has been shown over the first 40 round recommendations. We can see that the NICF, learned by directly learning to explore, favors for recommending more diverse items. The results verify that exploring users' interests can increase the recommendation diversity and enhancing diversity is also a means of improving exploration. 

\begin{figure*}
\includegraphics[width=4.4in]{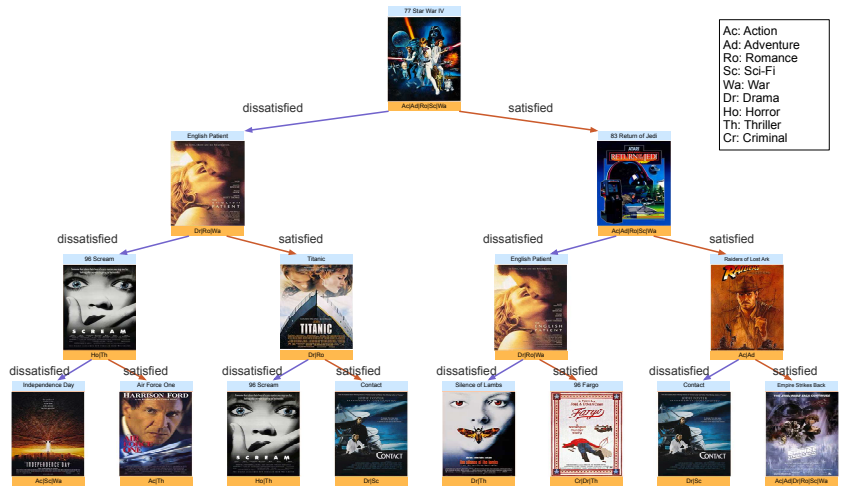}
    \caption{The sequential decision tree learned by NICF without using the genre information of movies.}
    \label{fig:rec_tree} 
\end{figure*}
\paragraph{The knowledge learned by NICF}
To gain a better insight into NICF, we take a close look at the exploration policy, \ie visualizing the sequential decision tree learned by NICF. Due to the space limitation, we only present the first four round recommendations on MovieLens (1M) dataset. As shown in the Figure \ref{fig:rec_tree}, without using the genre information, NICF can explore users' interests by recommending similar movies with some different topics if the user liked this movie, or changing the genre of the movies if the movie has been negative labeled, which indicates that NICF can effectively track users' interests and adapt its strategy to balance the exploration/exploitation on cold-start recommendations.

\section{Related Work}
We summarize the related literature: {\it traditional recommender system}, {\it interactive recommender system} and {\it meta-learning based recommender system} as follows.
\vspace{-0.1cm}
\paragraph{Traditional recommender system}  Being supervised by the history records and making recommendations with maximum estimated score have been the common practice in majority models, including \emph{factorization methods} ~\cite{rendle2010factorization,hoyer2004non,koren2009matrix} and different kinds of \emph{deep neural models}, such as multilayer perceptron~\cite{cheng2016wide,he2017neural}, denoising auto-encoders~\cite{wu2016collaborative}, convolutional neural network (CNN)~\cite{tang2018personalized}, recurrent neural network (RNN)~\cite{li2017neural,gu2020hierarchical}, memory network~\cite{chen2018sequential} and attention architectures~\cite{zhou2018deep,bai2019ctrec}. Based on the partially observed historical interactions, these existing models usually learn the user profile~\cite{zhou2018deep,gu2020hierarchical,chen2019semi,gu2016hlgps} and predict a customer's feedback by a learning function to maximize some well-defined evaluation metrics in the ranking, such as Precision and NDCG~\cite{clarke2008novelty}. However, most of them are myopic because the learned policies are greedy with estimating customers' feedback and unable to purposely explore users' interests for cold-start or warm-start users in a long term view. 
\vspace{-0.1cm}
\paragraph{Interactive recommender system} Interactive recommendation as a trend for the development of recommender systems has been widely studied in recent years. There are mainly two directions for the research: \textbf{(1)} contextual bandit; \textbf{(2)} reinforcement learning. \textbf{(1)} In contextual bandit, the main focus is on how to balance exploration and exploitation and achieving a bounded regret (\ie the performance gap between optimal recommendations and suggested recommendations) under worst cases. Hence, many contextual bandit based recommender systems have been developed for dealing with different recommendation tasks, such as news recommendation~\cite{li2010contextual}, diversify movie set~\cite{qin2014contextual}, collaborative filtering~\cite{wang2017factorization,zhao2013interactive}, online advertising\cite{zeng2016online} and e-commerce recommendation~\cite{wu2017returning}. However, they are usually intractable for non-linear models and potentially overly pessimistic about the recommendations. \textbf{(2)} Reinforcement learning is suitable to model the interactive recommender system. However, currently, there are still many difficulties in directly applying RL, such as the off-policy training~\cite{chen2019top,zou2020dynaq}, the off-policy evaluation~\cite{gilotte2018offline} and the large action spaces~\cite{dulac2015deep,zhao2018deep} and its topics are concentrated on optimizing the metrics with delayed attributes, such as diversity~\cite{zou2019diversity}, browsing depth~\cite{zou2019reinforcement}. As far as we know, we are the first work analyzing its usage on exploring users' interests for interactive collaborative filtering. 
\vspace{-0.1cm}
\paragraph{Meta-learning based recommender system}
Meta-learning, also called learning-to-learn, aims to train a model that can rapidly adapt to a new task with a few-shot of samples~\cite{finn2017model,koch2015siamese,santoro2016meta}, which is naturally suitable for solving the cold-start problem after collecting a handful of trial recommendations. For example, \citet{vartak2017meta} treated recommendation for one user as one task, and exploit learning to adopt neural networks across different tasks based on task information. ~\citet{lee2019melu} proposed to learn the initial weights of the neural networks for cold-start users based on Model-agnostic meta-learning (MAML)~\cite{finn2017model}. At the same time, ~\citet{pan2019warm} proposed a meta-learning based approach that learns to generate desirable initial embeddings for new ad IDs. However, all these methods ignored the performance on the support set, which also greatly influence the user engagement on the recommender system. Additionally, the full exploitation principle after few-shot trials inevitably led to the local optimal recommendations.
\section{Conclusions}
In this work, we study collaborative filtering in an interactive setting and focus on recommendations for cold-start users or warm-start users with taste drifting. To quickly catch up with users' interests, we propose to represent the exploration strategy with a multi-channel stacked self-attention neural network and learn it from the data. In our proposed method, the exploration strategy is encoded in the weights of the neural network, which are trained with efficient Q-learning by maximizing the cold-start or warm-start users' satisfaction in limited trials. The key insight is that the satisfying recommendations triggered by the exploration recommendation can be viewed as the delayed reward for the information gathered by exploration recommendation, and the exploration strategy that seamlessly integrates constructing the user profile into making accurate recommendations, therefore, can be directly optimized by maximizing the overall satisfaction with reinforcement learning. To verify its effectiveness, extensive experiments and analyses conducted on three benchmark collaborative filtering datasets have demonstrated the knowledge learned by our proposed method and its advantage over the state-of-the-art methods.

\section{Acknowledgement}
This research was supported by the Natural Sciences and Engineering Research Council (NSERC) of Canada. The authors gratefully appreciate all the anonymous reviewers for their valuable comments.
\bibliographystyle{ACM-Reference-Format}
\bibliography{reference} 
\end{document}